# Automatically Detecting Heterogeneous Bugs in High-Performance Computing Scientific Software

Matthew Davis
Georgia Institute of Technology
Atlanta, GA, USA
mdavis438@gatech.edu

Aakash Kulkarni
Oregon State University
Corvallis, OR, USA
kulkaraa@oregonstate.edu

Ziyan Chen
Georgia Institute of Technology
Atlanta, GA, USA
zchen910@gatech.edu

Yunhan Qiao
Oregon State University
Corvallis, OR, USA
qiaoy@oregonstate.edu

Christopher Terrazas
Oregon State University
Corvallis, OR, USA
terrazac@oregonstate.edu

Manish Motwani
Oregon State University
Corvallis, OR, USA
motwanim@oregonstate.edu

*Abstract*—Scientific advancements rely on high-performance computing (HPC) applications that model real-world phenomena through simulations. These applications process vast amounts of data on specialized accelerators (e.g., GPUs) using special libraries. Heterogeneous bugs occur in these applications when managing data movement across different platforms, such as CPUs and GPUs, leading to divergent behavior when using heterogeneous platforms compared to using only CPUs. Existing software testing techniques often fail to detect such bugs because either they do not account for platform-specific characteristics or target specific platforms. To address this problem, we present HeteroBugDetect, an automated approach to detect platform-dependent heterogeneous bugs in HPC scientific applications. HeteroBugDetect combines natural-language processing, off-target testing, custom fuzzing, and differential testing to provide an end-to-end solution for detecting platform-specific bugs in scientific applications. We evaluate HeteroBugDetect on LAMMPS, a molecular dynamics simulator, where it detected multiple heterogeneous bugs, enhancing its reliability across diverse HPC environments.

## I. Introduction

Scientific applications are essential tools for researchers across diverse domains. They simulate complex real-world phenomena, allowing scientists to test hypotheses and analyze intricate systems before undertaking costly and time-consuming physical experiments. These computationally intensive applications use accelerators like GPUs and specialized programming tools (e.g., OpenMP [1], CUDA [2], Kokkos[3]) to process large datasets efficiently. However, these specialized accelerators and tools can introduce a new class of errors known as heterogeneous bugs, which occur when data movement between heterogeneous devices is not properly managed.

For instance, errors may arise when a developer misuses APIs to transfer data between the CPU and GPU—where computations are performed—and then copies the results back to the CPU. These errors can only be identified by comparing application behavior on CPU-only platforms versus heterogeneous platforms combining CPUs and GPUs. HeteroBugDetect offers an end-to-end framework for automatically detecting such errors.

While there exists some techniques (e.g., HeteroFuzz [4], TitanFuzz [5], FuzzGPT [6], FuzzyFlow [7], HFuzz [8] etc.) for detecting heterogeneous bugs, they are typically tailored to specific applications or libraries and may not scale effectively for large, real-world scientific applications. HeteroBugDetect addresses this limitation by using the power of off-target testing [9], differential testing [10], and grammar-based fuzzing [11], augmented by state-of-the-art generative large language model (LLM) [12]. The primary goal of HeteroBugDetect is to improve the trustworthiness and reliability of scientific applications considering their inherent complexity and the diverse high-performance computing environments in which they operate.

The HeteroBugDetect has four key components that enhance the bug detection process in scientific applications. First, it uses an LLM to generate diverse inputs for the HPC system under test (SUT) by interpreting natural language descriptions of experiments, lowering the entry barrier for researchers unfamiliar with software testing. This improves the likelihood of detecting heterogeneous bugs that traditional methods might miss. Second, to address scalability, HeteroBugDetect extracts a standalone sub-system from the SUT using dynamic analysis and code transformation, streamlining the testing process by focusing on key code paths. Third, for generating test inputs that are more likely to expose bugs, it uses a grammar-based, kernel-sensitive fuzzing technique, mutating inputs based on changes in parallel executions and variables' properties that interact across heterogeneous platforms. Fourth, a differential driver runs platform-specific sub-systems on the same fuzzed inputs and compares outputs, flagging discrepancies that exceed tolerance as potential bugs. This tailored approach improves the efficiency and likelihood of detecting the heterogeneous bugs that may remain hidden when using traditional testing methodologies.

Since we could not find any existing defect benchmarks for heterogeneous bugs in scientific applications, we created our own benchmark by analyzing user-reported bug reports and possible mistakes that developers make in source code that lead to the creation of 20 heterogeneous bugs belonging to 7 categories in an open-source scientific application called LAMMPS [13]. Our evaluation of HeteroBugDetect shows that it can successfully detect 8 out of 20 bugs and additionally detect 2 *unknown* bugs, emphasizing its potential to enhance the robustness of scientific applications. We also perform an ablation study to assess the impact of off-target testing, grammar-based custom mutation, and kernel-sensitive fuzzing on the performance of HeteroBugDetect.

The main contributions of this paper are:
- HeteroBugDetect, an automated approach to detect heterogeneous bugs in HPC scientific applications.
- HeteroBench, the first benchmark of real-world heterogeneous bugs in a widely used scientific application.
- A prototype implementation of HeteroBugDetect supporting multiple backends and applications.

## II. BACKGROUND AND MOTIVATION

Scientific research has greatly advanced with high-performance computing (HPC) resources, including supercomputers, GPUs, and specialized frameworks like Kokkos [3] and Raja [14]. These technologies empower researchers across fields such as physics, chemistry, and biology to simulate complex phenomena and validate hypotheses with unprecedented accuracy. Central to this scientific renaissance are scientific applications—software programs specially designed to simulate real-world processes in a computational environment.

For example, LAMMPS (Large-scale Atomic/Molecular Massively Parallel Simulator)[13], a widely used open-source molecular dynamics software. Since 1995, hundreds of publications have relied on LAMMPS for simulating the behavior of atoms and molecules in diverse materials. LAMMPS is designed to run efficiently on heterogeneous platforms, including NVIDIA GPUs, Intel CPUs, and AMD GPUs, and using Kokkos for performance portability. However, the growing complexity of scientific applications such as LAMMPS introduces challenges in reliability, particularly with *heterogeneous bugs*. These bugs occur due to improper data synchronization between CPUs and accelerators, leading to subtle numerical inaccuracies or outright application crashes. Detecting and resolving such bugs is essential to maintaining trust in scientific simulations. Figure 1 shows the two plots illustrating a heterogeneous bug reported[1] by an experienced user when comparing the results of simulations performed in LAMMPS using different HPC environments. The plots show variation in the temperature (left plot) and volume (right plot) of an iron box over time steps that is heated to the temperature of 1500K quickly in a few steps and then heated more slowly. Depending on the interatomic forces set while initializing the simulation, the variation in temperature and volume of the box can vary. As shown, the user experiments with using the ReaxFF (Reactive Force Field) and Embedded Atom Method (EAM) potential, which are used to model interaction between atoms in metallic systems, such as pure metals and alloys. Further, the user also experiments running the same simulation (using ReaxFF potential) using two *different HPC environments*–with Kokkos library using GPU and without Kokkos library using CPU. When running the simulation without Kokkos, a linear expansion of the volume is observed. However, when using Kokkos, an abrupt rise in the volume and temperature occurs at step 1e+05 that *should not occur*. As domain scientists typically run their simulations on one HPC environment, they may never find out about such inconsistent results across different HPC environments, and can make decisions based on potentially misleading findings. While debugging the reported bug, developers found that this unexpected behavior occurs because a program variable *"wasn't being updated on GPUs correctly, and somehow this slipped through all the regression tests"*[2]. Thus, despite being heavily tested, the bug remained hidden from developers and was only discovered when a user compared the results of the same simulation across different HPC environments. This example demonstrates the need for a framework that targets detecting heterogeneous bugs and helps developers in enhancing their test suite.

Traditional testing methods may not be effective in detecting such heterogeneous bugs in large, real-world scientific applications because of their oversighted assumptions that the *applications take a minuscule amount of time (in ∼milliseconds) to execute and arbitrary mutations are likely to yield meaningful inputs*. Consequently, there arises a need for techniques that can detect and rectify heterogeneous bugs efficiently and accurately in scientific applications that use arbitrary hardware platforms. HeteroBugDetect can automatically detect such bugs by generating diverse simulation scripts (equivalent of test cases for scientific applications) and comparing the simulation results across user-selected HPC environments (e.g., executing LAMMPS using Kokkos with GPU and without using Kokkos on CPU).

## III. HETEROBUGDETECT APPROACH

Figure 2 shows the HeteroBugDetect architecture. As shown, HeteroBugDetect takes two inputs: the codebase of the HPC scientific software under test (SUT) and a natural language description of a supported simulation by the SUT. HeteroBugDetect reports potential heterogeneous bugs along with the bug-triggering test inputs (simulation scripts) to reproduce the bugs. The following sections describe the four key components of the HeteroBugDetect approach.

### A. Seed input generator

Scientific applications typically require domain scientists to create simulation scripts using specific commands and arguments to run simulations. Existing test-input generation techniques (e.g. EvoSuite [15], Randoop [16], Swami [17]) cannot generate such complex scripts to run simulations.

While developers may provide some example simulation scripts to demonstrate or test the use of their applications, these may not cover all possible features supported by the applications. Further, sometimes developers may only provide API reference documentation to assist in creating scripts. Thus, relying on developer-provided scripts would severely limit our technique's applicability. Alternatively, one could manually construct a well-formed grammar of the valid commands and arguments that can be used in simulation scripts, and use that grammar to generate more variants of scripts. However, unlike traditional software applications, the commands and arguments used in simulation scripts can have multiple inter-dependencies that often lead to producing syntactically valid but *semantically invalid* input scripts. Since our goal is to detect heterogeneous bugs, it is important to automatically generate simulation scripts that execute on at least one HPC environment without throwing any errors. Henceforth, we refer to a syntactically and semantically valid simulation scripts as valid simulation scripts.

---
[1]https://github.com/lammps/lammps/issues/3535

[2]https://github.com/lammps/lammps/pull/3541

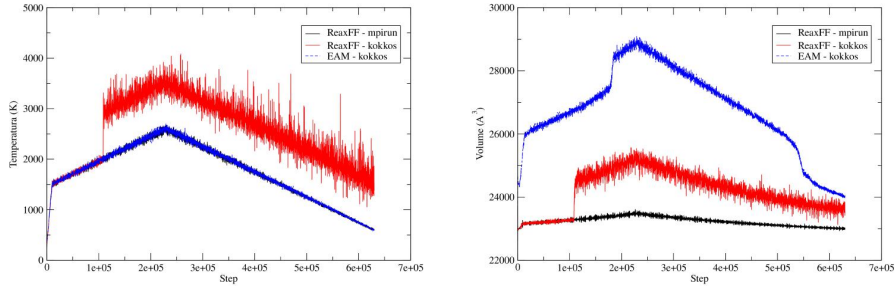

Fig. 1. Heterogeneous bug reported by a LAMMPS user. The plots show the variation in temperature (left) and volume (right) of a box heated in a simulation executed on LAMMPS using different HPC environments. The left plot shows that the variation in temperature is inconsistent across using the Kokkos+GPU and mpirun for the ReaxFF potential. The abrupt increase in the temperature occurs when using Kokkos because the data is not correctly updated on GPUs.

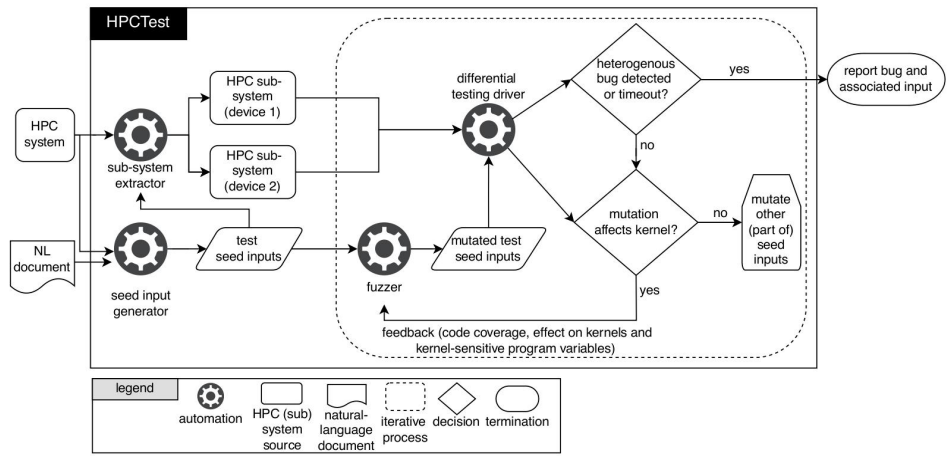

Fig. 2. The HeteroBugDetect architecture.

To address the challenge of automatically generating valid simulation scripts without relying on application developers, HeteroBugDetect leverages the power of modern natural language processing techniques using a generative large language model (LLM). Specifically, HeteroBugDetect uses OpenAI's *gpt-3.5-turbo* due to the model's top performance in general natural language processing tasks and within scientific research compared to other LLMs[18]. HeteroBugDetect generates diverse, valid simulation scripts for the HPC system-under-test (SUT) using the natural language descriptions of the simulations supported by the SUT ("seed input generator" in Figure 2) that are provided as input by the user.

The top of Figure 3 shows the generic prompt template that HeteroBugDetect uses for generating diverse and valid simulation scripts using a LLM. The bottom of the figure shows an instantiated prompt template to generate a simulation script for LAMMPS that uses the COMB potential from the examples listed in the LAMMPS manual [19]. As LLMs may not necessarily produce a valid simulation script in the first attempt, HeteroBugDetect uses the chain-of-thought [20] process to iteratively refine the output of an LLM by providing it the feedback obtained by parsing the result of executing the generated script on the SUT. Further, HeteroBugDetect attempts to produce many diverse scripts that may use different commands and arguments' values for the same prompt by varying relevant parameters of the LLM. We call this module

**Prompt template**: You are a helpful assistant who understands how to use <SUT name> for scientific simulation. Generate a simulation script that can be executed on <SUT name> to simulate <simulation description>. Do not provide any explanations and do not add any comments and blank lines in the script.

**Example prompt**: You are a helpful assistant who understands how to use LAMMPS for scientific simulation. Generate a simulation script that can be executed on LAMMPS to simulate models using the COMB potential. Do not provide any explanations and do not add any comments and blank lines in the script.
**Generated simulation script**:

```
1  units           metal
2  dimension       3
3  boundary        p p p
4  atom_style      charge
5  lattice         fcc 5.43
6  region          box block 0 10 0 10 0 10
7  create_box      1 box
8  create_atoms    1 box
9  mass            1 12.01
10 set             group all charge 0.0
11 pair_style      comb
12 pair_coeff      * * ffield.comb Cu
13 timestep        0.005
14 thermo          1
15 thermo_style    custom step temp pe ke etotal
                   press vol
16 thermo_modify   flush yes
17 velocity        all create 300 12345 dist
                   gaussian
18 fix             1 all nve
19 run             10
```

Fig. 3. The "Prompt template" shows the generic prompt used by HeteroBugDetect to generate diverse simulation scripts for HPC applications (SUT). The "Example prompt" shows the template instantiated for LAMMPS using the Comb example described in LAMMPS manual and the "Generated simulation script" shows a valid simulation script that executes successfully on LAMMPS without using Kokkos and GPUs.

---

**Algorithm 1:** Generating simulation scripts for the HPC system-under-test (SUT) using natural language description of the simulations supported by the SUT.

**Input:** natural language description of a simulation ($Sim_{desc}$), SUT name ($S_{name}$), SUT executable ($S_{exe}$), large langauge model ($L$), large langauge model parameters ($P$), $max\_attempts$
**Output:** scripts to run the input simulation on the system-under-test

1  **Procedure** generateSimulationScripts($Sim_{desc}$, $S_{name}$, $S_{exe}$)
2    $prompt\_t \leftarrow$ "You are a helpful assistant who understands how to use <SUT name> for scientific simulation. Generate a simulation script that can be executed on <SUT name> to simulate <simulation description>. Do not provide any explanations and do not add any comments and blank lines in the script."
3    $prompt \leftarrow$ initializePromptTemplate($prompt\_t$, $S_{name}$, $Sim_{desc}$)
4    **for** *combination c of LLM parameters P* **do**
5      $attempt \leftarrow 1$
6      **while** $attempt <= max\_attempts$ **do**
7        $LLM\_output \leftarrow$ generateScript($prompt$, $L$, $c$)
8        $error \leftarrow$ validateAndExtractError($LLM\_output$, $S_{exe}$)
9        **if** *error is null* **then**
10         saveScript($LLM\_output$, $attempt$)
11         break
12       **else**
13         $prompt \leftarrow$ concatenate($prompt$, $error$)
14       $attempt \leftarrow attempt + 1$

---

of HeteroBugDetect—that generates valid simulation scripts using an LLM—*Seed Input Generator (SIG)*. We refer to the generated valid simulation scripts as *seed inputs* because they form the initial set of seeds that are used as input in the subsequent modules of HeteroBugDetect.

Algorithm 1 explains the SIG module's seed input generation process within HeteroBugDetect. The SIG takes as input: (1) a natural language description of a simulation, (2) the name of the SUT, (3) the executable binary of the SUT, (4) a generative LLM, (5) a set of parameters for the LLM, and (6) the maximum number of attempts allowed to generate a valid simulation script. The SIG module first initializes the prompt template with the simulation description and name of the SUT (lines 2,3 in Algorithm 1). Next, for each combination of LLM parameter values (specified in a configuration file e.g., *temperature* values for using the *gpt-3.5-turbo* LLM), the SIG attempts to generate a valid simulation script up to a *max_attempts* times (lines 4–6 in Algoritrhm 1).

In each attempt, the SIG invokes the LLM using specific parameter values and provides an instantiated prompt. The SIG then uses regular expression-based parsing to parse the LLM response and store the generated script (line 7 in Algorithm 1). Next, the SIG validates the stored script by executing it on the SUT and parsing any error messages if they occur. Typically, invalid scripts throw errors instantly. However, to prevent generating scripts that can take a very long time to run, the SIG uses a timeout (specified in seconds) to bound execution time (line 8 in Algorithm 1). If the generated script does not throw any errors or does not timeout, the SIG saves the script along with recording the metadata including the attempt in which that script was generated and the LLM's parameter values. The SIG module then continues to generate more diverse scripts for the same description using different parameter values (lines 9–11 in Algorithm 1). However, if the generated script throws errors or timeouts, the SIG concatenates the error messages thrown by the SUT (or "modify the script to reduce the execution time" in case of timeout) to the prompt used for generating that script and re-attempts to generate a valid script for $max\_attempts$ number of times (lines 12–14 in Algorithm 1). The output of the SIG module results in multiple valid scripts for executing diverse simulations in the SUT.

### B. Sub-system extractor

As scientific applications provide a wide variety of features for the domain scientists to experiment with, this makes such applications' code base more complex compared to traditional software systems. Furthermore, as executing a single simulation can range from a few seconds to hours, and simulation scripts typically contain commands and arguments with multiple interdependencies, using traditional fuzzing methods that assume *it takes a minuscule amount of time in the order of milliseconds to execute a target application and arbitrary mutations are likely to yield meaningful inputs* **does not work well**. The reason is because instrumenting and exploring the entire code base of scientific applications to perform effective coverage-guided fuzzing takes a significant amount of time.

To address this challenge, HeteroBugDetect uses off-target testing [9] to carve out a small standalone executable func-tionalities from large code-bases. For this, HeteroBugDetect employs dynamic program analysis to detect and extract the *sub-systems* from the system-under-test (SUT) that: (1) can be executed as an standalone application, (2) are specific to the given example simulation, and (3) are specific to the HPC environments used for executing the simulation ("sub-system extractor" in Figure 2).

The sub-system-extractor (SSX) module takes as input: (1) the source code of the SUT, (2) the seed inputs generated by the SIG module (recall Section III-A) for a given example simulation, and (3) at least two HPC environments in which the SUT should be executed. The output of this module are the two sub-systems that only contain the source code required to run the simulation on the two input HPC environments.

The SSX module works as follows. First, it generates a compilation database for the target SUT that includes the dependencies specific to HPC environments. For example, to generate a LAMMPS sub system that executes using Kokkos library on GPUs, SSX generates a compilation database that contains all the build commands executed when building LAMMPS that will use Kokkos enabled with GPUs.

Next, SSX generates a trace log of all the functions that are invoked by all the simulation scripts of an example simulation. For this, SSX parses the abstract syntax tree (AST) of the SUT source code to detect all the functions and transforms their AST to enable logging mechanisms so that when the instrumented application runs, it will generate a trace file containing all functions executed. For this, SSX adds a logging statement at the start of each function that invokes SSX's built-in function called *instrumentor* using arguments: (class name, function name, and function return type). Whenever the instrumentor is invoked, it first checks if the function invoked is already logged using a hashset of strings. If the function is not logged, SSX logs it by adding it to the hashset. The output of this step is a single trace log that lists all the functions that are *required* to execute all the scripts of an example simulation on a specific HPC environment. SSX stubs all non-required functions using the trace log (e.g., using std::terminate() in C++ or pass in Python), rendering them unreachable and simplifying the fuzzing process. This step is crucial to streamline the testing process and ensure that the most pertinent code paths in the SUT are thoroughly scrutinized.

### C. Grammar-Based Kernel-Sensitive Fuzzer

Traditional fuzzing works by randomly mutating seed inputs and executing them on system-under-test (SUT) to identify security vulnerabilities where developers miss validating inputs. While using random fuzzing[21][22] is effective in detecting security vulnerabilities in many real-world systems, our experiments show that it does not work well in detecting heterogeneous bugs, which require executing the SUT using valid simulation scripts on multiple HPC environments.

To address this challenge, HeteroBugDetect uses grammar-based kernel-sensitive fuzzing ("fuzzer" in Figure 2), where it mutates the seed inputs considering the syntax of commands

```
1  "create_box": {
2      "arg_1": [1, 2, 3],
3      "arg_2": ["No change"]
4  },
5  "create_atoms":{
6      "arg_1": [1, 2, 3],
7      "style_arg": {
8          "box": [],
9          "region": ["No change"],
10         "random": {
11             "arg_1": ["integer"],
12             "arg_2": ["integer"],
13             "arg_3": ["No change"]
14         }
15     }
16 },
17 "mass": {
18     "arg_1": [1, 2, 3, "C", "*"],
19     "arg_2": ["float"]
20 },
```

Fig. 4. Partial view of the manually-constructed grammar in JSON format used by HeteroBugDetect to generate diverse simulation scripts for LAMMPS.

and their arguments supported by the SUT. Furthermore, HeteroBugDetect uses the insight that mutating commands in the simulation script that invoke parallel computations and properties of *kernel-sensitive variables*, which are the internal program variables (e.g., Kokkos views) accessed/modified by any kernel (code offloaded to a device for parallel execution) in the SUT is more likely to reveal heterogeneous bugs.

To enable grammar-based mutations, two of the authors manually constructed the grammar of valid inputs supported by the SUT using the documentation manual provided by the developers. This grammar is represented in the JSON format and it contains various commands, their arguments and allowed values supported by the SUT. As a command may take multiple optional arguments and values, and may also have interdependence on other commands, the grammar is carefully constructed to capture these nuances that guide HeteroBugDetect to generate semantically valid inputs. **This is a one-time effort** and it took the authors a couple of days to construct and refine the grammar for a complex application such as LAMMPS.

Figure 4 shows a partial view of the grammar constructed for LAMMPS that shows arguments and values for the `create_box`, `create_atoms`, and `mass` commands. As shown, the grammar uses keywords "No change" to direct the fuzzer that it should not mutate that argument's value. Similarly, the grammar uses keyword "style_arg" to direct the fuzzer to randomly select one or more entries from the choices. For example, while mutating `create_atoms` command, the value of first argument ("arg_1") must be selected from the list `[1, 2, 3]` and the next argument(s) should be constructed using one or more choices randomly selected from the "style_arg" entries. This results in valid mutations of the simulation scripts used as seed inputs.

While grammar-based fuzzing is effective in producing semantically valid inputs for the SUT, it may still not be effective in detecting heterogeneous bugs because it selects all commands in the simulation script for mutation uniformly, whereas only some commands may actually affect parallel computations in the SUT. Thus, HeteroBugDetect enhances the traditional grammar-based fuzzing by dynamically detecting and mutating those commands that affect parallel computations. For this, HeteroBugDetect monitors the change in parallel computations and their associated program variables' properties when executing the SUT using mutated simulation script on the two HPC environments, and adjusts the selection probabilities of lines in the script based on the observed changes.

Specifically, for each fuzzing iteration in which a single line is mutated, HeteroBugDetect computes a *signal* that varies from $-1$ to $+1$ and captures the overall changes observed in the following six metrics each of which indicates how the mutated line affected parallel computations.

**(1) Memory Leak Fraction (ML):** The fraction of kernel-sensitive variables whose allocated memory space is not completely freed after executing the fuzzer-generated input. This is computed by measuring the difference between the sum of space allocated and de-allocated for each variable throughout the program execution and computing the fraction of variables showing memory leaks over all variables. The value of this metric varies from 0 to 1.

**(2) Parallel Loop Count (PL):** The number of parallel for loops triggered during the execution of a fuzzer-generated input script. The value of this metric is a positive integer $\geq 0$.

**(3) Parallel Scan Count (PS):** The number of parallel scan events (e.g. prefix sum, prefix max) triggered during the execution of a fuzzer-generated input script. The value of this metric is a positive integer $\geq 0$.

**(4) Parallel Reduction Count (PR):** The number of parallel reduce events (e.g. sum, max) triggered during the execution of a fuzzer-generated input script. The value of this metric is a positive integer $\geq 0$.

**(5) Fence Execution Count (FE):** The number of fence events (barriers encountered between execution of multiple kernels) triggered during the execution of a fuzzer-generated input script. The value of this metric is a positive integer $\geq 0$.

**(6) Deep Copy Count (DC):** The number of deep copy events (copying data of some program variable across heterogeneous devices e.g., CPU to GPU) triggered during the execution of a fuzzer-generated input script. The value of this metric is a positive integer $\geq 0$.

At the start of fuzzing process, all lines in the seed input are equally likely to be selected for mutation and therefore, have a selection probability of $1/n$, where $n$ is the total number of lines in the seed input. When running the fuzzer for the first time, HeteroBugDetect computes and records the values of five metrics based on the execution of the fuzzer-generated output script. In subsequent iterations, the tool recomputes these metrics and calculates the differences between the current and previous iteration's metric values. The overall difference (*signal*), is computed as the average of the normalized differences across all five metrics that belong to the range $[-1, +1]$. To ensure the signal remains within a consistent range, the average score is divided by 2, yielding a final signal value that varies between $[-0.5, +0.5]$. The greater the signal, the higher the likelihood of finding heterogeneous bugs in the SUT by mutating *the same command* in the fuzzed scripts. The positive value of the signal indicates that the line mutated increases parallel computations while the negative signal indicates that the line mutated decreases parallel computations and therefore should be less prioritized when selecting lines for mutation during fuzzing. The magnitude of the signal indicates the amount by which the selection probabilities of the lines should be adjusted. HeteroBugDetect adjusts the selection probability of the line mutated using the following formula:

$$probability = probability * (1 + signal)$$

Thus, the selection probability of a line is increased if it improves parallel computations and decreased if it reduces them. The selection probabilities of all other lines are adjusted such that the sum of all selection probabilities adds to 1.

### D. Differential Driver

Finally, traditional coverage-guided fuzzing cannot be directly used to detect heterogeneous bugs in scientific applications because executing the application on two HPC environments would spawn two different processes, while for the coverage analysis, the fuzzer needs to analyze the code covered while executing the application on two different HPC environments *within the same process*. Further, since the goal of this study is not to detect crashes or security vulnerabilities, and scientific simulations may produce slightly different results because of randomness and floating-point precision used in underlying hardware, we cannot just `diff` the logs produced using different HPC environments and report bugs when logs differ as this would result in many false positives.

To address these challenges, HeteroBugDetect uses a differential driver component ("differential testing driver" in Figure 2) that reformulates traditional fuzzing to report heterogeneous bugs instead of crashes.

To execute the HPC system-under-test (SUT) on two different HPC environments within the same process, differential driver compiles the SUT as a static library and invokes the library's functions to launch the simulation twice using a different set of arguments specifying the simulation script and the HPC environment parameters. This allows the fuzzer to measure the code coverage on the SUT considering the execution on both HPC environments in a single process.

To filter out false-positives that arise due to minor differences in the simulation results produced using different HPC environments, the differential driver provides various error norms (L1 [Manhattan], L2 [Euclidean], and Max [Infinity] norms) that are typically used in scientific simulations with an error threshold that can be set by the user to make HeteroBugDetect report a bug only if the differences observed using error norm are greater than threshold. If HeteroBugDetect generates a script whose simulation results across two HPC environments differ more than the threshold, it reports that observation as a heterogeneous bug along with the input script that causes divergent outputs.

| bug category | bug IDs | # of bugs |
|---|---|---|
| Incorrect synchronization between host and device variables | 1, 4, 6, 11, 15, 16, 18, 19 | 8 |
| Missing synchronization between host and device variables | 7, 8, 10, 12, 13, 17 | 6 |
| Accessing memory in device space from host | 2 | 1 |
| Missed copying variable data from host to device or vice versa | 5 | 1 |
| Incorrect copying of variables from host to device or vice versa | 20 | 1 |
| Use of stale data | 9, 14 | 2 |
| Concurrent modification to shared variable from host and device | 3 | 1 |
| total | | 20 |

Fig. 5. Types of heterogeneous bugs and their associated bug IDS and count in the HeteroBench benchmark constructed for evaluating HeteroBugDetect.

## IV. EVALUATION

This section describes the dataset, metrics, experiment procedures, and the evaluation results for HeteroBugDetect.

### A. Dataset

To the best of our knowledge, no standard benchmark exists for heterogeneous bugs in real-world scientific applications. To address this gap, we introduce HeteroBench, a defect benchmark comprising 20 bugs categorized into seven distinct types. The benchmark was constructed by mining GitHub issues reported by LAMMPS users related to heterogeneous bugs (see Section II) and engaging with developers of scientific applications such as LAMMPS, Kokkos, and ArborX [23]. We also analyzed discussions on public forums for these applications to identify potential root causes of heterogeneous bugs. Based on these analyses, we defined seven bug categories and curated HeteroBench, which includes 20 bugs distributed across all categories (Figure 5). Of these, four are user-reported bugs, while 16 were injected into the LAMMPS codebase by the authors. For each bug, the benchmark provides a buggy version of the code, a fixed version, and a simulation script to reproduce the bug.

### B. Evaluation Metrics

To evaluate the performance of HeteroBugDetect, we use the following metrics:

- *Line Coverage (LC)*: This measures the number of lines of code exercised by inputs generated during fuzzing. Higher line coverage indicates broader exploration of the codebase, which increases the likelihood of uncovering heterogeneous bugs.
- *Unique bug count (UBC)*: The total number of distinct heterogeneous bugs detected during fuzzing. A bug is considered unique if its associated source code line in the error message generated by the bug-causing input differs from those of other bugs.
- *Benchmark performance (BP)*: This measures the number of heterogeneous bugs detected by a fuzzing technique when tested against the HeteroBench bug benchmark. This metric evaluates the effectiveness of a fuzzing approach in identifying different categories of heterogeneous bugs.
- *Coverage Per Input (CPI)*: This measures the percentage of total line coverage achieved per input during fuzzing. It is calculated by dividing the total line coverage percentage by the total number of inputs. Higher CPI indicates that each input contributes significantly to code coverage efficiency. We use this metric because the likelihood of detecting heterogeneous bugs, the primary goal of this study, is higher in code sections that go beyond input validation, which is better captured by this metric.

### C. Implementation and Experiment Procedure

We implemented HeteroBugDetect's SIG module using GPT-3.5 [24] to generate initial seed inputs. The SSX module leverages the LLVM compiler infrastructure [25] to detect and instrument functions in the target application, enabling trace generation and function stubbing. The Fuzzer module is built on AFL++ v4.10c [26], incorporating grammar-based fuzzing capabilities. For kernel-sensitive profiling, we extended Kokkos Tools [27], a framework designed for efficient profiling, debugging, and tuning of HPC applications without requiring recompilation. Specifically, HeteroBugDetect utilizes the *MemoryEvents* and *KernelLogger* tools from Kokkos Tools to develop *Director*, a custom profiling tool that generates logs used to compute the metrics described in Section III-C. The differential driver module, implemented in C++, employs error norms (L1, L2, Max) to detect divergences in simulation results across different HPC environments.

We evaluated HeteroBugDetect on LAMMPS using 58 examples from [19], creating natural language descriptions for each. For the 40 examples whose seed inputs were successfully generated, HeteroBugDetect's SSX module extracted the corresponding subsystems, taking an average of 104 seconds ($\sim$ 2 minutes) per example. Subsequently, for each example, HeteroBugDetect's fuzzer module was executed with 10 random seeds, with a 30-minute timeout per seed. The fuzzer module ran for 7 hours to detect heterogeneous bugs, employing three different HeteroBugDetect configurations as part of the ablation study. Overall, the experiments required approximately two weeks of wall-clock time, with an additional week allocated for analysis and result computation. All experiments were executed on an Ubuntu 24.04 server equipped with 2 Nvidia A40 GPUs, with each fuzzing run assigned to a single GPU.

### D. Results

This section describes our evaluation results in terms of the three research questions we ask.

*1) RQ1: How effective is HeteroBugDetect in detecting heterogeneous bugs?:* Figure 6 shows the HeteroBugDetect performance in terms of the line coverage (LC), coverage per valid input (CPI), unique bug count (UBC), and benchmark performance (BP) for each of the 10 runs when executing it to detect heterogeneous bugs in LAMMPS.

As shown, HeteroBugDetect produces only syntactically valid inputs, generating an average of 960 valid inputs across 10 runs. Considering LC, which evaluates the efficiency of fuzzing strategy in generating meaningful inputs that explore distinct execution paths–HeteroBugDetect achieves coverage values ranging from 12.9% to 14.5% across 10 runs, demonstrating consistent performance. On average, HeteroBugDetect achieves a line coverage of 13.68%, indicating its ability to consistently generate inputs that effectively probe the program's execution paths.

| Seed  | # Valid | # Invalid | LC(%) | CPI   | UBC | BP     |
|-------|---------|-----------|-------|-------|-----|--------|
| 118   | 1004    | 0         | 14.50 | 0.014 | 2   |        |
| 1985  | 932     | 0         | 14.50 | 0.016 | 2   | 3, 10  |
| 5729  | 889     | 0         | 14.10 | 0.016 | 1   |        |
| 12287 | 1023    | 0         | 14.30 | 0.014 | 1   | 14, 15 |
| 22136 | 986     | 0         | 14.00 | 0.014 | 1   |        |
| 24719 | 950     | 0         | 13.50 | 0.014 | 1   |        |
| 24857 | 920     | 0         | 12.90 | 0.014 | 1   | 16, 18 |
| 26284 | 992     | 0         | 12.60 | 0.013 | 1   |        |
| 26941 | 953     | 0         | 13.40 | 0.014 | 1   | 19, 20 |
| 26973 | 949     | 0         | 13.00 | 0.014 | 1   |        |
| avg   | 960     | 0         | 13.68 | 0.014 | —   | —      |
| total | —       | —         | —     | —     | 2   | 8      |

Fig. 6. Coverage per input (CPI), unique bug count (UBC), and benchmark performance (BP) of HeteroBugDetect across 10 runs while testing LAMMPS using 40 examples

Considering coverage per input (CPI), that quantifies the coverage achieved per input and evaluates the efficiency of fuzzing strategy in generating meaningful inputs that explore distinct execution paths, HeteroBugDetect achieves CPI ranging from 0.013 to 0.014 across 10 runs, demonstrating consistent performance. On average, HeteroBugDetect achieves a CPI of 0.014, indicating its ability to consistently generate inputs that effectively probe the program's execution paths.

Considering unique bug count (UBC), HeteroBugDetect uncovered two previously unknown heterogeneous bugs in LAMMPS by generating inputs that revealed divergent behaviors when executed across different HPC environments. The top panel of Figure 7 shows a HeteroBugDetect-generated script using the *Airebo* example, while the bottom panel highlights differences in the output logs generated by executing this script on LAMMPS across two HPC configurations: LAMMPS with Kokkos and CUDA on a GPU, and LAMMPS without Kokkos on a CPU. The input script caused the simulation to crash after step 10 on the CPU, whereas on the GPU, it crashed after step 20. The error messages, originating source files, and spikes in Temp, TotEng, and Press differ significantly between the two environments. The second unknown bug was identified using a HeteroBugDetect-generated script for the *Comb* example. In this case, LAMMPS executed with Kokkos on a GPU crashed with the error: *Lost atoms: original 4000 current 3999 (src/thermo.cpp:491)*. However, the same script executed successfully and produced a valid log when run without Kokkos on a CPU. These findings have been reported to the LAMMPS developers, and we are currently awaiting their response.

Finally, in benchmark performance (BP), HeteroBugDetect successfully detected 8 (40%) (bug ID# 3, 10, 14, 15, 16, 18, 19, and 20) out of 20 defects in the HeteroBench benchmark. The detected heterogeneous defects belong to five distinct cat-

```
1   units metal
2   dimension 3
3   boundary p p p
4   atom_style full
5   atom_modify map array sort 0 0.0
6   read_data polyethylene.data
7   pair_style airebo 3.0
8   pair_coeff * * CH.airebo C H
9   bond_style harmonic
10  bond_coeff 1 2.59 4.12
11  angle_style harmonic
12  angle_coeff 1 2.68 4.24
13  thermo_style custom step temp etotal press density
14  thermo 10
15  neighbor 2.0 bin
16  neigh_modify delay 0 every 8 check yes
17  fix 1 all nve
18  run 20
```

```
--- log.gpu.txt
+++ log.cpu.txt
...
Unit style       : metal
Current step     : 0
Time step        : 0.001
-Per MPI rank memory allocation (min/avg/max) = 8.383 | 8.383 | 8.383 Mbytes
-Step Temp         TotEng        Press         Density
-0    0            587.40916     12957.368     0.014556699
-10   271999.75    718.6126      32305.355     0.014556699
-ERROR on proc 0: Bond atoms 1 3 missing on proc 0 at step 16 (src/
     ntopo_bond_all.cpp:59)
+Per MPI rank memory allocation (min/avg/max) = 8.809 | 8.809 | 8.809 Mbytes
+Step Temp         TotEng        Press         Density
+0    0            587.40916     12957.368     0.014556699
+10   271999.75    718.6126      32305.355     0.014556699
+20   2.2184837e+11 1.5463623e+09 -9.9300032e+10 0.014556699
+ERROR on proc 0: Atoms have moved too far apart (-2000.5306571379392) for
     minimum image (src/domain.cpp:986)
```

Fig. 7. HeteroBugDetect-generated script (top) for LAMMPS that produces divergent results (bottom) when executed using different HPC environments.

egories: (1) incorrect synchronization between host and device variables, (2) concurrent modifications to shared variables by host and device, (3) missing synchronization between host and device variables, (4) use of stale data, and (5) incorrect copying of variables between host and device. These results highlight HeteroBugDetect's effectiveness in identifying diverse heterogeneous bugs, showcasing its potential as a valuable tool for improving the reliability of HPC scientific applications.

> HeteroBugDetect demonstrates robust performance by consistently generating syntactically valid inputs with an average line coverage of 13.68%, average coverage per valid input of 0.014, uncovering two previously unknown heterogeneous bugs in LAMMPS, and detecting 8 out of 20 defects across five distinct categories in the HeteroBench benchmark.

*2) RQ2: How does HeteroBugDetect's perform compare to baseline?:* To our knowledge, no existing technique detects heterogeneous bugs in real-world scientific applications that require complex inputs for simulations. Therefore, we use random fuzzing as a baseline to evaluate HeteroBugDetect. Random fuzzing, implemented with AFL++ v4.10c [26], generates diverse inputs through mutation operations like bit-flipping, byte insertion, and altering integers in various ways. By comparing HeteroBugDetect to random fuzzing, we establish a reference point for evaluating its performance. We executed random fuzzing using the seed inputs generated by HeteroBugDetect's SIG module for the 40 examples on the entire LAMMPS codebase using the same 10 randomly generated seeds and the same timeout of 30 min per seed as we used while executing HeteroBugDetect.

| technique | # Valid Inputs | # Invalid | LC (%) | UBC | BP |
|---|---|---|---|---|---|
| AFL++ (Random) | 7 | 11 | 3.80 | 0 | 0 |
| HeteroBugDetect | 958 | 0 | 13.97 | 1 | 8 |

Fig. 8. Comparison of Random fuzzing and HeteroBugDetect performance average over 6 runs while testing LAMMPS using 40 examples.

In our evaluation, we observed that random fuzzing struggled under the same time constraints as HeteroBugDetect, failing to generate scripts in 4 out of 10 runs due to timeouts. This issue arises from the inefficiency of processing a larger codebase and numerous seed inputs each of which can take a few seconds, within the allocated time. In contrast, HeteroBugDetect isolates smaller subsystems, reducing complexity and enabling consistent performance within the same time budget. This highlights the limitations of traditional testing methods for large-scale scientific applications.

To ensure a fair and meaningful comparison, we evaluated our technique against random fuzzing using only the 6 runs where random fuzzing successfully generated results. Figure 8 shows the results of the evaluation in terms of the metrics defined in Section IV-B, averaged over 6 runs. As shown, there is a significant difference in performance between the random fuzzing technique and HeteroBugDetect. The random fuzzing technique generated only 7 valid inputs and covered just 3.8% of the lines of code (LC), with no unique bugs (UBC) or benchmark performance (BP) identified. In contrast, HeteroBugDetect generated 958 valid inputs, covering 13.97% lines of the code, and successfully identified 1 unique bug along with 8 defects from the HeteroBench benchmark. These findings highlight the superior effectiveness of HeteroBugDetect in generating a much larger number of valid inputs, covering a higher percentage of the code, and identifying more defects. The results suggest that HeteroBugDetect's targeted fuzzing approach, likely through more efficient seed input generation and coverage-guided techniques, significantly outperforms random fuzzing.

> The HeteroBugDetect surpasses the baseline by delivering higher input efficiency and successfully identifying bugs, making it a more effective tool for scientific software testing and heterogeneous bug detection.

*3) RQ3: How do different components of HeteroBugDetect contribute to its performance?:* To answer this research question, we analyze the contribution of four key components of HeteroBugDetect: the Seed Input Generator (SIG), the Subsystem Extractor (SSX), the grammar-based fuzzing (Custom), and the metrics to measure the effect on parallel computations to guide the fuzzer.

**SIG effectiveness**: To evaluate the effectiveness of HeteroBugDetect's seed input generation, we measured how many example simulations SIG successfully produced valid scripts for LAMMPS. The more examples SIG can generate valid scripts for, the more effective HeteroBugDetect is in generating diverse test cases. Out of the 58 examples for which LAMMPS can be executed using Kokkos on GPUs and sequentially on CPUs, SIG successfully generated valid input scripts for 40 (69%) examples, averaging three attempts per example. These scripts were subsequently used in HeteroBugDetect's fuzzing component. Of the 40 examples, 27 (67.5%) required no modifications, while the remaining 13 (32.5%) needed minor manual adjustments. These adjustments included specifying paths to domain-specific data files (e.g., for simulations using the EIM potential, the required data file was provided by developers[3] or correcting the order of commands in the generated scripts. The number of distinct scripts generated for each of these 40 examples ranged from 1 to 15, with an average of 4 scripts per example. In total, SIG generated 170 unique seed inputs for these 40 examples.

**SSX effectiveness**: To evaluate the contribution of SSX in HeteroBugDetect's performance, we measure how much it reduces the codebase, as this directly impacts fuzzing performance. By reducing the number of lines, functions, and branches, SSX simplifies the fuzzing process and improves its efficiency. Across 40 examples, the results demonstrate substantial reductions: lines of code are reduced by an average of 81.99%, branches by 82.07%, and functions by 36.48%. These reductions highlight the capability of SSX to eliminate irrelevant portions of the code, enabling HeteroBugDetect to focus on the most critical areas, which can significantly enhance both performance and bug discovery.

To further assess SSX's impact on HeteroBugDetect's performance, we ran HeteroBugDetect on the full codebase and subsystems from the 40 examples using 10 runs and the same timeout per run. As shown in the top two rows of Figure 9, when using the full codebase, HeteroBugDetect generated 12.7 valid inputs, covered 6.71% of the code, with a CPI of 0.0055, detecting 2 unique bugs and achieving a BP of 2. In contrast, when using subsystems, HeteroBugDetect generated 959.8 valid inputs, covered 13.68% of the code, and had a CPI of 0.0143. It detected the same 2 unique bugs but identified 8 additional defects from the benchmark. These results highlight the effectiveness of focusing on subsystems for fuzzing, which leads to more valid inputs, better code coverage, and enhanced defect detection compared to testing the entire codebase.

**Grammar-based fuzzing and kernel-sensitive metrics effectiveness**: For this, we created two variants of HeteroBugDetect: HeteroBugDetect$_{Random}$ and HeteroBugDetect$_{Custom}$. The bottom two rows in the Figure 9 show the performance of HeteroBugDetect when using random fuzzing and when using grammar-based custom fuzzing without using the metrics.

Comparing HeteroBugDetect with HeteroBugDetect$_{Random}$ reveals that while using random fuzzing generated higher number of valid inputs (2273.9) and covered slightly more lines of code (14.34), it showed a very low CPI (0.004) indicating that most of the inputs generated are similar covering similar lines of code. It also did not identify any unique bugs (UBC), though it detected 3 defects in the HeterBench benchmark (BP). In contrast, HeteroBugDetect produced fewer valid inputs (959.8) while achieving a slightly lower line coverage (13.68%) but a significantly higher CPI (0.0143). It successfully identified 2 unique bugs and detected 8 BP issues. This suggests that while random fuzzing can generate a larger number of inputs, using grammar-based kernel-sensitive fuzzing provides more targeted and efficient defect detection and achieving similar coverage despite generating fewer inputs.

---
[3]https://github.com/lammps/lammps/blob/develop/potentials/ffield.eim

| technique | #Valid | #Invalid | LC(%) | CPI | UBC | BP |
|---|---|---|---|---|---|---|
| HeteroBugDetect (Full Codebase) | 12.7 | 0 | 6.71 | 0.0055 | 2 | 2 |
| HeteroBugDetect | 959.8 | 0 | 13.68 | 0.0143 | 2 | 8 |
| HeteroBugDetect$_{Random}$ | 2273.9 | 1460.1 | 14.34 | 0.0040 | 0 | 3 |
| HeteroBugDetect$_{Custom}$ | 963.9 | 0 | 13.43 | 0.0137 | 2 | 8 |

Fig. 9. Ablation study results showing the effect of using subsystem extractor, grammar-based fuzzing, and kernel-sensitive metrics on HeteroBugDetect's performance. The results are averaged across 10 runs.

When comparing HeteroBugDetect$_{Custom}$ with HeteroBugDetect to identify the effect of using metrics, the results show that HeteroBugDetect achieves better input efficiency, as evidenced by its higher CPI of 0.0143 compared to 0.0137 for $HeteroBugDetect_{custom}$. This improvement in CPI indicates that HeteroBugDetect, which uses specialized metrics to prioritize mutations, enhances the efficiency of input generation without compromising on defect detection. Both techniques generated a similar number of valid inputs (963.9 for HeteroBugDetect$_{Custom}$ vs. 959.8 for HeteroBugDetect) and achieved nearly identical line coverage (13.43% for HeteroBugDetect$_{Custom}$ vs. 13.68% for HeteroBugDetect). Furthermore, both approaches successfully identified 2 unique bugs and detected 8 BP issues. These findings show that using kernel-sensitive metrics further improves the fuzzing efficiency while maintaining strong performance in bug detection.

> Each component of HeteroBugDetect enhances performance. The SIG produced 170 diverse seed inputs across 40 examples, SSX reduced the codebase by over 80%, leading to more valid inputs and higher code coverage. Grammar-based kernel-sensitive fuzzing achieved better CPI, identified more defects, and uncovered unique bugs missed by random fuzzing.

## V. Related Work

**Hardware-Focused Strategies.** In the lens of HPC systems, testing can be difficult due to multiple device environments. HeteroFuzz [4] focuses on an approach that speeds up the traditional fuzzing process considerably. While the speed to fuzz a heterogeneous program is improved, the scope of testing an HPC application is limited to specific C and C++ domains and its use of static analysis. Unlike static analysis, which struggles with abstractions like C++ templates, HeteroBugDetect uses dynamic analysis for a granular view of HPC applications. Similarly, HFuzz [8] speeds up the fuzzing process significantly by exploiting hardware optimizations such as strategic injections within the host and kernel environments and parallelization of the fuzzing process. Our strategy uses hardware accelerators (GPUs) within its full pipeline for efficient computation, but with a focus on a complete end-to-end solution for HPC scientific applications.

**Piecemeal Strategies.** FuzzyFlow [7] frames the HPC environment as a graph problem and speeds up the testing process by reducing the input space by extracting a min-cut from a data flow graph. However, this approach will not capture bugs that rely solely on probalistic fuzzing. To mitigate this, we also divide the HPC environment, but rather than graph cuts, we extract sub-systems from an entire HPC application to increase the diversity of our testing. As with LAMMPS [13] and other HPC scientific applications, the code bases can be large and have many interconnected subsystems in their simulations. HeteroBugDetect ensures that bugs within the subsystems of an entire HPC system, even with low probablity from a fuzzer, can still be detected. With this approach, we are able to increase coverage of these types of bugs while not sacrificing performance.

**LLMs and Fuzzing.** TitanFuzz [5] and FuzzGPT [6] are closest to our approach in utilizing LLMs within the fuzzing process. TitanFuzz [5] effectively uses LLMs to fuzz deep learning libraries, but is limited in its scope to only this domain. While deep learning libraries utilize multple backend architectures, they are only a subset within the entire HPC system ecosystem. FuzzGPT [6] capitalizes on prior bug knowledge for improved in-context learning for the LLM by manually creating a bug-inducing dataset from Github repositories. In our approach, we utilize the chain-of-thought process for LLMs only as an automated approach to create valid seed files. This strategy has the advantage of eliminating the need to mine repositories and reliance on domain experts.

## VI. Discussion and Threats to Validity

We recognize that HeteroBugDetect's dynamic analysis, focused on subsystem extraction from example inputs, may leave parts of the application untested, limiting overall coverage. While we explored a static-analysis-based solution to address this, it proved impractical for heavily templated C++ code common in HPC applications, as template instantiation occurs only at build time. Thus, we opted for a dynamic approach, which, while not exhaustive, identifies issues with high accuracy. Expanding the range of natural language example inputs and leveraging improved LLMs can further enhance its effectiveness, underscoring HeteroBugDetect's practicality and extensibility.

To address threats to internal validity, we carefully tested our implementation and have made our code and data publicly available for transparency and reproducibility. HeteroBench, though currently limited to 20 bugs from LAMMPS, is the first benchmark of heterogeneous bugs in real-world applications. Constructing such benchmarks for diverse scientific applications is challenging, requiring deep domain expertise and knowledge of HPC environments. To address this, we incorporated cross-application insights from Kokkos and ArborX to ensure generalizable bug categories. We are actively expanding HeteroBench with additional applications and will release it as an open-source resource to foster broader community contributions.

## VII. CONTRIBUTIONS

We presented HeteroBugDetect, a dynamic-analysis-based approach for detecting heterogeneous bugs in large-scale HPC scientific applications. By combining natural language processing, subsystem extraction, fuzzing, and differential testing, HeteroBugDetect overcomes the limitations of traditional methods and effectively uncovers diverse bugs, including previously unknown issues, in LAMMPS. We also introduced HeteroBench, the first benchmark for heterogeneous bugs, to support reproducibility and community engagement. By releasing our code and data publicly, we aim to foster collaboration and further advancements in HPC testing. This work sets a foundation for improving the reliability of HPC software and advancing heterogeneous bug detection.

## REFERENCES

[1] L. Dagum and R. Menon, "Openmp: an industry standard api for shared-memory programming," *IEEE Computational Science and Engineering*, vol. 5, no. 1, pp. 46–55, 1998.

[2] D. Luebke, "Cuda: Scalable parallel programming for high-performance scientific computing," in *2008 5th IEEE International Symposium on Biomedical Imaging: From Nano to Macro*. Piscataway, NJ, USA: IEEE (Institute of Electrical and Electronics Engineers), 2008, pp. 836–838.

[3] C. R. Trott, D. Lebrun-Grandié, D. Arndt, J. Ciesko, V. Dang, N. Ellingwood, R. Gayatri, E. Harvey, D. S. Hollman, D. Ibanez, N. Liber, J. Madsen, J. Miles, D. Poliakoff, A. Powell, S. Rajamanickam, M. Simberg, D. Sunderland, B. Turcksin, and J. Wilke, "Kokkos 3: Programming model extensions for the exascale era," in *IEEE Transactions on Parallel and Distributed Systems*, vol. 33, no. 4, 2022, pp. 805–817.

[4] Q. Zhang, J. Wang, and M. Kim, "Heterofuzz: Fuzz testing to detect platform dependent divergence for heterogeneous applications," in *Proceedings of the 29th ACM Joint Meeting on European Software Engineering Conference and Symposium on the Foundations of Software Engineering*, ser. ESEC/FSE 2021. New York, NY, USA: Association for Computing Machinery, 2021, p. 242–254. [Online]. Available: https://doi.org/10.1145/3468264.3468610

[5] Y. Deng, C. S. Xia, H. Peng, C. Yang, and L. Zhang, "Large language models are zero-shot fuzzers: Fuzzing deep-learning libraries via large language models," in *Proceedings of the 32nd ACM SIGSOFT International Symposium on Software Testing and Analysis*, ser. ISSTA 2023. New York, NY, USA: Association for Computing Machinery, 2023, p. 423–435. [Online]. Available: https://doi.org/10.1145/3597926.3598067

[6] Y. Deng, C. S. Xia, C. Yang, S. D. Zhang, S. Yang, and L. Zhang, "Large language models are edge-case generators: Crafting unusual programs for fuzzing deep learning libraries," in *Proceedings of the IEEE/ACM 46th International Conference on Software Engineering*, ser. ICSE '24. New York, NY, USA: Association for Computing Machinery, 2024. [Online]. Available: https://doi.org/10.1145/3597503.3623343

[7] P. Schaad, T. Schneider, T. Ben-Nun, A. Calotoiu, A. N. Ziogas, and T. Hoefler, "Fuzzyflow: Leveraging dataflow to find and squash program optimization bugs," in *Proceedings of the International Conference for High Performance Computing, Networking, Storage and Analysis*, ser. SC '23. New York, NY, USA: Association for Computing Machinery, 2023. [Online]. Available: https://doi.org/10.1145/3581784.3613214

[8] J. Wang, Q. Zhang, H. Rong, G. H. Xu, and M. Kim, "Leveraging hardware probes and optimizations for accelerating fuzz testing of heterogeneous applications," in *Proceedings of the 31st ACM Joint European Software Engineering Conference and Symposium on the Foundations of Software Engineering*, ser. ESEC/FSE 2023. New York, NY, USA: Association for Computing Machinery, 2023, p. 1101–1113. [Online]. Available: https://doi.org/10.1145/3611643.3616318

[9] T. Kuchta and B. Zator, "Auto off-target: Enabling thorough and scalable testing for complex software systems," in *Proceedings of the 37th IEEE/ACM International Conference on Automated Software Engineering*, ser. ASE '22. New York, NY, USA: Association for Computing Machinery, 2023. [Online]. Available: https://doi.org/10.1145/3551349.3556915

[10] W. M. McKeeman, "Differential testing for software," *Digital Technical Journal*, vol. 10, no. 1, pp. 100–107, 1998.

[11] P. Godefroid, A. Kiezun, and M. Y. Levin, "Grammar-based whitebox fuzzing," *SIGPLAN Not.*, vol. 43, no. 6, p. 206–215, jun 2008. [Online]. Available: https://doi.org/10.1145/1379022.1375607

[12] F. Barreto, L. Moharkar, M. Shirodkar, V. Sarode, S. Gonsalves, and A. Johns, "Generative artificial intelligence: Opportunities and challenges of large language models," in *International Conference on Intelligent Computing and Networking*, Springer. Singapore: Springer Nature Singapore, 2023, pp. 545–553.

[13] A. P. Thompson, H. M. Aktulga, R. Berger, D. S. Bolintineanu, W. M. Brown, P. S. Crozier, P. J. in 't Veld, A. Kohlmeyer, S. G. Moore, T. D. Nguyen, R. Shan, M. J. Stevens, J. Tranchida, C. Trott, and S. J. Plimpton, "Lammps - a flexible simulation tool for particle-based materials modeling at the atomic, meso, and continuum scales," *Computer Physics Communications*, vol. 271, p. 108171, 2022. [Online]. Available: https://www.sciencedirect.com/science/article/pii/S0010465521002836

[14] D. A. Beckingsale, J. Burmark, R. Hornung, H. Jones, W. Killian, A. J. Kunen, O. Pearce, P. Robinson, B. S. Ryujin, and T. R. Scogland, "Raja: Portable performance for large-scale scientific applications," in *2019 IEEE/ACM International Workshop on Performance, Portability and Productivity in HPC (P3HPC)*. Piscataway, NJ, USA: IEEE, 2019, pp. 71–81.

[15] G. Fraser and A. Arcuri, "Evosuite: automatic test suite generation for object-oriented software," in *Proceedings of the 19th ACM SIGSOFT Symposium and the 13th European Conference on Foundations of Software Engineering*, ser. ESEC/FSE '11. New York, NY, USA: Association for Computing Machinery, 2011, p. 416–419. [Online]. Available: https://doi.org/10.1145/2025113.2025179

[16] C. Pacheco and M. D. Ernst, "Randoop: feedback-directed random testing for java," in *Companion to the 22nd ACM SIGPLAN Conference on Object-Oriented Programming Systems and Applications Companion*, ser. OOPSLA '07. New York, NY, USA: Association for Computing Machinery, 2007, p. 815–816. [Online]. Available: https://doi.org/10.1145/1297846.1297902

[17] M. Motwani and Y. Brun, "Automatically generating precise oracles from structured natural language specifications," in *Proceedings of the 41st International Conference on Software Engineering*, ser. ICSE '19. Montreal, Quebec, Canada: IEEE Press, 2019, p. 188–199. [Online]. Available: https://doi.org/10.1109/ICSE.2019.00035

[18] W. X. Zhao, K. Zhou, J. Li, T. Tang, X. Wang, Y. Hou, Y. Min, B. Zhang, J. Zhang, Z. Dong, Y. Du, C. Yang, Y. Chen, Z. Chen, J. Jiang, R. Ren, Y. Li, X. Tang, Z. Liu, P. Liu, J.-Y. Nie, and J.-R. Wen, "A survey of large language models," 2024. [Online]. Available: https://arxiv.org/abs/2303.18223

[19] LAMMPS Documentation Team, "Examples - lammps molecular dynamics simulator," n.d., accessed: 2024-12-09. [Online]. Available: https://docs.lammps.org/Examples.html

[20] J. Wei, X. Wang, D. Schuurmans, M. Bosma, brian ichter, F. Xia, E. Chi, Q. V. Le, and D. Zhou, "Chain-of-thought prompting elicits reasoning in large language models," in *Advances in Neural Information Processing Systems*, S. Koyejo, S. Mohamed, A. Agarwal, D. Belgrave, K. Cho, and A. Oh, Eds., vol. 35. Red Hook, NY, USA: Curran Associates, Inc., 2022, pp. 24824–24837. [Online]. Available: https://proceedings.neurips.cc/paper_files/paper/2022/file/9d5609613524ecf4f15af0f7b31abca4-Paper-Conference.pdf

[21] C. Chen, B. Cui, J. Ma, R. Wu, J. Guo, and W. Liu, "A systematic review of fuzzing techniques," *Computers & Security*, vol. 75, pp. 118–137, 2018. [Online]. Available: https://www.sciencedirect.com/science/article/pii/S0167404818300658

[22] X. Zhao, H. Qu, J. Xu, X. Li, W. Lv, and G.-G. Wang, "A systematic review of fuzzing," *Soft Comput.*, vol. 28, no. 6, pp. 5493–5522, oct 2023. [Online]. Available: https://doi.org/10.1007/s00500-023-09306-2

[23] D. Lebrun-Grandié, A. Prokopenko, B. Turcksin, and S. R. Slattery, "Arborx: A performance portable geometric search library," *ACM Trans. Math. Softw.*, vol. 47, no. 1, Dec. 2020. [Online]. Available: https://doi.org/10.1145/3412558

[24] J. Ye, X. Chen, N. Xu, C. Zu, Z. Shao, S. Liu, Y. Cui, Z. Zhou, C. Gong, Y. Shen, J. Zhou, S. Chen, T. Gui, Q. Zhang, and X. Huang, "A comprehensive capability analysis of gpt-3 and gpt-3.5 series models," 2023. [Online]. Available: https://arxiv.org/abs/2303.10420

[25] C. Lattner, "Introduction to the llvm compiler infrastructure," in *Itanium conference and expo*, 2006.

[26] A. Fioraldi, D. Maier, H. Eißfeldt, and M. Heuse, "AFL++ : Combining incremental steps of fuzzing research," in *14th USENIX Workshop on Offensive Technologies (WOOT 20)*. USENIX Association, Aug. 2020. [Online]. Available: https://www.usenix.org/conference/woot20/presentation/fioraldi

[27] H. C. Edwards and C. R. Trott, "Kokkos: Enabling performance portability across manycore architectures," in *2013 Extreme Scaling Workshop (xsw 2013)*. IEEE, 2013, pp. 18–24.